\documentclass[]{iopart}

\expandafter\let\csname equation*\endcsname\relax
\expandafter\let\csname endequation*\endcsname\relax

\usepackage[unicode]{hyperref}
\usepackage[all]{hypcap}
\usepackage{amssymb,amsmath,verbatim,mathtools,needspace,enumitem,etoolbox,graphicx,physics,microtype,pifont,url}
\usepackage[usenames,dvipsnames]{xcolor}
\usepackage[T1]{fontenc}
\usepackage[utf8]{inputenc}
\usepackage[normalem]{ulem}
\definecolor{linkcolor}{rgb}{0.0,0.3,0.5}
\hypersetup{colorlinks=true,linkcolor=linkcolor,citecolor=linkcolor,filecolor=linkcolor,urlcolor=linkcolor}
\usepackage{orcidlink}
\usepackage{lipsum}
\newcommand{\ssim}{\mathchar"5218\relax\,}

\newcommand{\bham}{{School of Physics and Astronomy \& Institute for Gravitational Wave Astronomy, University of Birmingham, Birmingham, B15 2TT, UK}}
\newcommand{\milan}{{Dipartimento di Fisica ``G. Occhialini'', Universit\'a degli Studi di Milano-Bicocca, Piazza della Scienza 3, 20126 Milano, Italy}}
\newcommand{\infn}{{INFN, Sezione di Milano-Bicocca, Piazza della Scienza 3, 20126 Milano, Italy}}

\begin{document}

\begin{center}
\title[D.~Gerosa and M.~Bellotti]{Quick recipes for gravitational-wave selection effects}
\end{center}

\author{
Davide Gerosa$^{1,2,3}$ \orcidlink{0000-0002-0933-3579},
Malvina Bellotti$^{1}$ \orcidlink{0009-0003-1050-9273}
}
\vspace{0.1cm}
\address{$^{1}$~\milan}
\address{$^{2}$~\infn}
\address{$^{3}$~\bham}

\ead{\href{mailto:davide.gerosa@unimib.it}{\rm davide.gerosa@unimib.it}}

\setcounter{footnote}{0}

\begin{abstract}
Accurate modeling of selection effects is a key ingredient to the success of gravitational-wave astronomy. The detection probability plays a crucial role in  both statistical population studies, where it enters the hierarchical Bayesian likelihood, and astrophysical modeling, where it is used to convert predictions from population-synthesis codes into observable distributions. We review the most commonly used approximations, extend them, and present some recipes for a straightforward implementation. These include a closed-form expression capturing both multiple detectors and noise realizations written in terms of the so-called Marcum $Q$-function and a ready-to-use mapping between  signal-to-noise ratio thresholds and false-alarm rates from state-of-the-art detection pipelines. The bias introduced by approximating the matched filter signal-to-noise ratio with the optimal signal-to-noise ratio is not symmetric: sources that are nominally below threshold are more likely to be detected than sources above threshold are to be missed. Using both analytical considerations and software injections in detection pipelines, we confirm that including noise realizations when estimating the selection function introduces an average variation of a few~\%. This effect is most relevant for large catalogs and specific subpopulations of sources at the edge of detectability (e.g. high redshifts).
\end{abstract}

\section{Introduction}

Gravitation-wave (GW)  surveys are affected by selection biases. GW selection effects are relatively clean to model compared to those of conventional (i.e. electromagnetic) 
astronomy because, unlike photons, GWs travel 
 unaffected across the Universe from emission to detection. %
That said, our detectors are not equally sensitive to compact binaries with different parameters  (masses, spins, distance, inclination, etc.), which implies their  observational coverage is not uniform. 
Selection-effect modeling is crucial in GW population studies~\cite{2019MNRAS.486.1086M,2022hgwa.bookE..45V} and it is not an exaggeration to say that accurately estimating the probability of detection ---commonly referred to as $p_{\rm det}$--- is a key ingredient to the success of GW astronomy as a whole. Indeed, it was shown \cite{2023MNRAS.526.3495T} that modeling errors in the selection function will be (and perhaps already are) the leading limiting factor in our astrophysical inference, or at least that which scales more severely with the number of observed events.

The modern and more accurate approach to estimating selection biases is that of performing software injections using the same pipelines that are used for detection~\cite{2023PhRvX..13a1048A}. While accurate, this procedure is computationally expensive, requiring reweighting schemes \cite{2018CQGra..35n5009T}, calibration factors~\cite{calibratedVT}, 
and an effective number of samples that is at least a factor of a few greater than the number of events in the catalog~\cite{2019RNAAS...3...66F}. In practice, many astrophysical predictions in the field  relies on (semi-)analytical approximations to $p_{\rm det}$. Seminal work in this direction was presented by Finn and Chernoff \cite{1993PhRvD..47.2198F}, who approximated the detection statistics using the optimal signal-to-noise ratio (SNR) and factored out the dependence on the binary extrinsic parameters. %
Their approach is still widely adopted, with SNR thresholds
$\rho_{\rm t}$ ranging from $\ssim 8$ to $\ssim 12$ \cite{2018LRR....21....3A}. %
More recent advances %
include those by Essick \cite{2023PhRvD.108d3011E}, who presented a semi-analytical model of GW detectability capturing the finite size of template banks, with all the associated complexities. Additional attempts leveraging both analytical~\cite{2021ApJ...922..258V} and machine-learning~\cite{2020PhRvD.102j3020G,2022ApJ...927...76T} techniques have also been explored. One can also calibrate SNR thresholds a posteriori, i.e. using the events that are considered detected \cite{2024PhRvD.109f3013M}.

In this paper, we review and extend some of the most commonly used approximations to GW selection effects, %
notably including noise realizations (Sec.~\ref{secdect}). For a single detector, we further develop the marginalization over the extrinsic parameters first presented in Ref.~\cite{1993PhRvD..47.2198F}. For multiple detectors, we %
 show that thresholding the matched filter SNR 
 results in an expression for $p_{\rm det}$ that can be written down in closed form using special functions. In short, it is sufficient to substitute a sharp step function with a so-called Marcum $Q$-function~\cite{Marcum1948AST,Shnidman1989TheCO,2013arXiv1311.0681G}. %
We apply our findings to both controlled distributions and LIGO/Virgo injections  (Sec.~\ref{secapp}). Thresholding the matched filter SNR instead of the optimal SNR results in values of $p_{\rm det}$ that are systematically higher, and thus merger rates that are systematically lower. While this effect is of a few \%, its impact grows dramatically with the size of the GW catalog \cite{2023MNRAS.526.3495T} and disproportionally affects those specific regions of the parameters space where sources are close to the detection %
horizon \cite{2020ApJ...891L..31F,2023PhRvD.107j1302M,2024ApJ...962..169E}.
Our expressions are straightforward to implement
 and
can be used to quickly post-process large samples of simulated sources such as the outputs of astrophysical population-synthesis codes  (Sec.~\ref{secend}).  To facilitate the exploitation of our findings, we also describe a straightforward implementation of the Marcum $Q$-function for the Python programming language (\ref{python}).

\section{SNR thresholds}
\label{secdect}

We organize our calculation %
in three steps of increasing complexity.  
In what follows, the symbol $\theta$ collectively denotes the intrinsic parameters of a compact binary (e.g. masses, spins) as well as the distance to the sources, while $\xi$ indicates the remaining extrinsic parameters (inclination, sky location, and polarization angle). 

\subsection{Single detector \& optimal SNR}

Let us first consider the case of a single detector. The optimal SNR is given by
\begin{equation}
\label{snrintegral}
\rho_{\rm opt} = 2 \sqrt{\int_0^\infty \frac{|h(f)|^2}{S(f)} \dd f} \,.
\end{equation}
where $f$ indicates frequency, $h(f)$ is the GW strain, and $S(f)$ is the one-sided power spectral density of the detector. %
The word ``optimal'' has sometimes been used in the literature (including by some of the authors~\cite{2020PhRvD.102j3020G}) to indicate the SNR of an optimally oriented source. In this paper, we refer to the more common meaning of optimality with respect to noise realizations. 

A common approximation \cite{1993PhRvD..47.2198F} is to take $\rho_{\rm opt} $ as a detection statistics, i.e. assume that a source is detectable if the optimal SNR is greater than a given threshold $\rho_{\rm t}$. That is, we write
\begin{equation}
p_{\rm det}(\theta,\xi) = \mathcal{I}[\rho_{\rm opt} (\theta,\xi) > \rho_{\rm t}]\,,
\end{equation}
where $\mathcal{I}$ is an indicator function equal to one if the condition inside the brackets is true and zero otherwise.\footnote{Note that $p_{\rm det}(\theta,\xi)\in[0,1]$ is a probability and not a probability density, i.e. it does not integrate to unity over $\theta$ and $\xi$. One should more carefully write $p({\rm det}| \theta,\xi)$.} 
For a given distribution of extrinsic parameters $p(\xi)$, one can compute (with an abuse of notation)
\begin{equation}
\label{pdetthetadef}
p_{\rm det}(\theta) = \int p_{\rm det}(\theta, \xi) p(\xi) \dd\xi \,.
\end{equation}
For a single detector, the optimal SNR factorizes as follows~\cite{1993PhRvD..47.2198F,2020PhRvD.102j3020G}
\begin{equation}
\label{factoromega}
\rho_{\rm opt} (\theta,\xi) = \omega(\xi) \rho_{\rm max}(\theta)\,,
\end{equation}
where $0\leq \omega\leq 1$ is a projection parameter and  $\rho_{\rm max}(\theta)$ is the optimal SNR of an optimally oriented source (i.e. a binary with face-on inclination located overhead the detector). In particular, one has $\xi=\{\iota,\vartheta,\phi,\psi\}$ and 
\begin{align}
\omega \,=\; &\Bigg\{\left(\frac{1+\cos^2\iota}{2} \right)^2 \bigg[ \frac{1}{2}\left (1+\cos^2\vartheta \right )\cos 2\phi \cos 2\psi  
 - \cos \vartheta\sin 2\phi \sin 2\psi\bigg]^2 
 \notag \\
& + 
\cos^2\iota \bigg[ \frac{1}{2}\left (1+\cos^2\vartheta \right )
 \cos 2\phi \sin 2\psi + 
\cos\vartheta\sin 2\phi \cos 2\psi  \bigg]^2  \Bigg\}^{1/2}
\,,
\end{align}
where $\iota$ is the binary inclination, $\vartheta$ and $\phi$ are a polar and an azimuthal angle for the sky location, and $\psi$ is the polarization angle. Note that the factorization of Eq.~(\ref{factoromega}) breaks down for precessing sources because the inclination of the orbital plane depends on the emitted frequency. That said, this effect is likely to be mild because current GW observations cover $\lesssim 1$ precession cycle.

With this factorization, the integral in Eq.~(\ref{pdetthetadef}) becomes \cite{1993PhRvD..47.2198F}
\begin{equation}
\label{pdetthetaeasy}
p_{\rm det}(\theta) = \!\int\! \mathcal{I}[\omega \rho_{\rm max} (\theta,\xi) > \rho_{\rm t}] p(\omega) \dd\omega  =\! \int_{\rho_{\rm t}/ \rho_{\rm max}(\theta)}^\infty\!\!\!\!\!\!\!\!\!\!\!\!\!\!\!\!p(\omega) \dd\omega\,. 
\end{equation}
Assuming that the distribution $p(\xi)$ of the extrinsic parameters is known, this implies that selection effects can be estimated by evaluating the complementary cumulative distribution function of $\omega$  \cite{1993PhRvD..47.2198F}
\begin{equation}
\label{cumulative}
p(\omega>\omega_0) = \int_{\omega_0}^\infty p(\omega') \dd\omega'\,.
\end{equation}
The simplicity of this approximation is appealing: estimating selection effects reduces to evaluating a single waveform $h(f)$ for an optimally oriented source with intrinsic parameters $\theta$, calculating $\rho_{\rm max}$ from  Eq.~(\ref{snrintegral}), and evaluating the universal function $p(>\omega)$  at $\omega= \rho_{\rm t}/ \rho_{\rm max}$. 

\begin{figure}
\centering
\includegraphics[width=0.65\columnwidth]{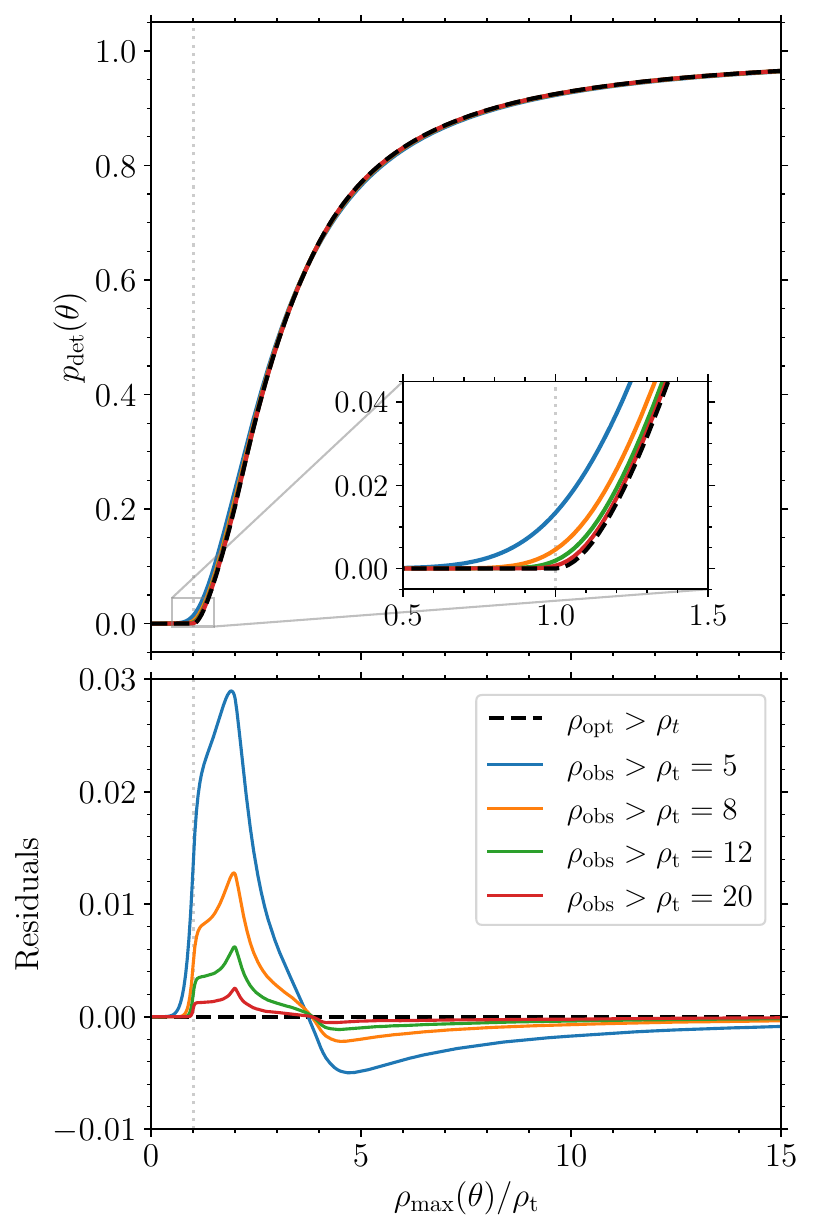}
\caption{Detection probability $p_{\rm det}(\theta)$ averaged over the extrinsic parameters as a function of the optimal, single-detector SNR $\rho_{\rm max}(\theta)$ of an optimally oriented source with intrinsic parameters $\theta$. The black-dashed line is obtained by thresholding the optimal SNR. Colored solid lines are obtained by thresholding the matched filter SNR with various thresholds $\rho_{\rm t}$. The top panel shows the detection probability itself, while the bottom panel shows the difference between $p_{\rm det}$ obtained using the matched filter SNR as a detection statistics and $p_{\rm det}$ obtained using the optimal SNR instead. The vertical dotted lines indicate $\rho_{\rm max}(\theta)=\rho_{\rm t}$.} %
\label{onedetector}
\end{figure}

Figure~\ref{onedetector} shows the outcome of this procedure assuming that sources  are distributed isotropically in the sky (i.e. we take uniform distributions in $\cos\iota$, $\cos\vartheta$, $\phi$, and $\psi$). To facilitate comparison with the rest of the paper, we show the averaged detectability $p_{\rm det}(\theta)$ as a function of $\rho_{\rm max}/ \rho_{\rm t}$ instead of its inverse, even though the latter  enters Eq.~(\ref{pdetthetaeasy}) directly. For instance, the black curve in the top panel of Fig.~\ref{onedetector} indicates that at least $\ssim 80\%$ of the binaries with a given set of intrinsic parameters $\theta$ will be detectable if at least one of these sources has an SNR that is $\ssim 5$ times above the detection threshold.

\subsection{Single detector \& matched filter SNR}

Because noise realizations, any specific GW source will not be observed with SNR $\rho_{\rm opt}$ but rather with some other value $\rho_{\rm obs}$. In the standard matched-filtering approach to GW detection, $\rho_{\rm obs}$ is given by the filtered signal (made of both GWs and noise) normalized by its own root-mean-square; see Refs. \cite{Maggiore:2007ulw,2011gwpa.book.....C}.
One can thus include the effect of noise realizations in the estimate of $p_{\rm det}$ by thresholding the observed SNR $\rho_{\rm obs}$ instead of $\rho_{\rm opt}$ and computing the expectation value over noise realizations $n$, i.e.
\begin{equation}
\label{intovernoise}
p_{\rm det}(\theta,\xi) = \int \mathcal{I}[\rho_{\rm obs}(n, \theta,\xi) > \rho_{\rm t}] \, p(n)\, \dd n\,.
\end{equation}

 Assuming the noise in the detector is Gaussian and stationary, the observed SNR is distributed normally around the optimal SNR with unit variance (see  Refs.~\cite{Maggiore:2007ulw,2011gwpa.book.....C} but  also Ref.~\cite{2023PhRvD.108d3011E} for caveats). Because of this property, computing $\rho_{\rm opt}$ reduces to evaluating in Eq.~(\ref{snrintegral}) and adding a variance of one. One has
\begin{equation}
\label{gauss}
p(\rho_{\rm obs}) = \frac{1}{\sqrt{2\pi}} \exp\left[ -  \frac{(\rho_{\rm obs} - \rho_{\rm opt})^2}{2} \right]\,,
\end{equation}
and thus  \cite{2019PASA...36...10T}
\begin{align}
p_{\rm det}(\theta,\xi) &=\frac{1}{\sqrt{2\pi}} \int_{\rho_{\rm t}}^{\infty}  \exp\left\{ -  \frac{[\rho_{\rm obs} - \rho_{\rm opt}( \theta,\xi)]^2}{2} \right\} \dd\rho_{\rm obs} 
\notag \\ & = \frac{1}{2}\left\{ 1 + \erf\left[\frac{\rho_{\rm opt}(\theta,\xi) - \rho_{\rm t}}{\sqrt{2}} \right] \right\}\,,
\end{align}
where $\erf$ is the error function. Marginalizing over the extrinsic parameters as above yields %
\begin{align}
p_{\rm det}(\theta) = \frac{1}{2}\left\{ 1 + \int \erf\left[\frac{\omega\rho_{\rm max}(\theta) - \rho_{\rm t}}{\sqrt{2}} \right] p(\omega) \dd\omega \right\}\,.
\end{align}
Our results are shown in Fig.~\ref{onedetector} assuming isotropic sources.  Note that, unlike Eq.~(\ref{pdetthetaeasy}), thresholding the observed SNR does not result in a universal function of $\rho_{\rm max}(\theta)/\rho_{\rm t}$ but rather a one-parameter family of functions, where the additional parameter is $\rho_{t}$ itself. %

In practice, introducing the SNR variance due to noise realizations causes variations in $p_{\rm det}$ that are of $\mathcal{O}(1\%)$. The effect decreases as $\rho_{\rm t}$ increases: if the threshold is large, it is less likely that introducing a variance of one might turn a detectable event into a non-detectable event, or vice versa. Broadly speaking, we find that the GW detectability computed by thresholding the matched filter SNR is larger (smaller) than that obtained by thresholding the optimal SNR when signals are weak (loud). From Fig.~\ref{onedetector}, the transition between these two behaviors takes place at $\rho_{\rm max}\simeq  4 \rho_{\rm t}$. The largest deviations are found at SNRs $\rho_{\rm max}\simeq  2 \rho_{\rm t}$.

The luminosity distance $d_L$ of astrophysical objects in the nearby Universe is distributed geometrically, $p(d_L)\propto  d_L^{-2}$, which implies that $\rho_{\rm max}\propto 1/d_L$  is distributed as $p(\rho_{\rm max}) \propto {\rho_{\rm max}^{-4}}$~\cite{2011CQGra..28l5023S} (but note this will not be true for third-generation detectors~\cite{2016PhRvD..94l1501V}). This implies that the left part of the curve in Fig.~\ref{onedetector} where $\rho_{\rm t}\simeq \rho_{\rm max}$ has a disproportionally larger impact on the overall population of detected sources. Therefore, thresholding the optimal SNR instead of the observed SNR has the net effect of underestimating $p_{\rm det}$ (i.e. the residuals in Fig.~\ref{onedetector} are positive) and thus overestimating the astrophysical merger rate.

One caveat here is that we did not truncate the Gaussian distribution in Eq.~(\ref{gauss}) to impose $\rho_{\rm obs}\geq 0$. This is appropriate as long as the threshold value is much greater than the SNR variance, i.e. $\rho_{\rm t}\gg 1$. %

\subsection{Multiple detectors \& matched filter SNR}
\label{multdetmatched}
Let us now consider a network of multiple detectors. The network SNR is the root sum square of the individual SNRs, i.e. 
\begin{equation}
\rho_{\rm obs} (\{n\}, \theta,\xi)  = \sqrt{ \sum_i^{N}  \rho^2_{\rm obs,i} (n_i, \theta,\xi) }\,,
\end{equation}
where $N$ is the number of interferometers.
Each of the $\rho_{\rm obs,i}$'s is distributed normally around optimal values  $ \rho_{\rm opt, i}$ with unit variance. The square root of the sum of Gaussian variates with different means and unit variances is distributed according to the non-central $\chi$ distribution (which is also known as the generalized Rayleigh distribution)~\cite{Park1961MomentsOT}; see also Ref.~\cite{2023PhRvD.108d3011E}. The probability density function of $\rho_{\rm obs}$ is  
\begin{align}
\label{prhoobsnet}
p(\rho_{\rm obs} ) &=  {\rho}_{\rm opt} \left( \frac{\rho_{\rm obs}}{{\rho}_{\rm opt}} \right)^{
 N/2}  \exp\left(-\frac{\rho_{\rm obs}^2 + {\rho}_{\rm opt}^2}{2}\right)
\,
 I_{N/2 -1}({\rho}_{\rm opt}  \rho_{\rm obs} ) \,,
\end{align}
where $I_\nu(z)$ is the modified Bessel function of the first kind and order $\nu$ \cite{1972hmfw.book.....A}.
The parameter $\rho_{\rm opt}$ in Eq.~(\ref{prhoobsnet}) is root sum square of the individual optimal SNRs calculated as in Eq.~(\ref{snrintegral}), i.e. %
\begin{equation}
\rho_{\rm opt} (\theta,\xi)  = \sqrt{ \sum_i^{N}  \rho^2_{\rm opt,i} (\theta,\xi) }\,.
\label{rhobaropt}
\end{equation} 
Note how the detection probability in Eq.~(\ref{prhoobsnet}) only depends on the combined quantity $\rho_{\rm opt} $ and not on how this SNR is distributed among the various instruments in the network.%

We can now threshold the matched filter SNR $\rho_{\rm obs}$ and compute the expectation value over noise realizations as in Eq.~(\ref{intovernoise}). We find this can also be written down using special functions. In particular, one has 
\begin{equation}
\label{Qn2}
p_{\rm det}(\theta,\xi) = Q_{N/2}\big[\rho_{\rm opt} (\theta,\xi), \rho_{\rm t} \big]\,,
\end{equation}
where  %
\begin{equation} 
\label{Qdef}
Q_{\nu }(a,b)=  a^{1-\nu}\int _{b}^{\infty }x^{\nu }\exp \left(-{\frac {x^{2}+a^{2}}{2}}\right)I_{\nu -1}(ax)\,\dd x
\end{equation}
 is the %
 generalized Marcum $Q$-function of order $\nu$~\cite{Marcum1948AST,Shnidman1989TheCO,2013arXiv1311.0681G}, which is used in the field of digital communications (but see e.g. Refs.~\cite{2015PhRvD..91l4062G,2012LRR....15....4J} for some previous appearances in GW astronomy). In words, the generalized Marcum $Q$-function is the complementary cumulative distribution function of the non-central $\chi$ distribution. A two-line Python code to evaluate Eq.~(\ref{Qdef})  is described in \ref{python} and made available at Ref.~\cite{repo}.
 
\begin{figure}\centering
\includegraphics[width=0.65\columnwidth]{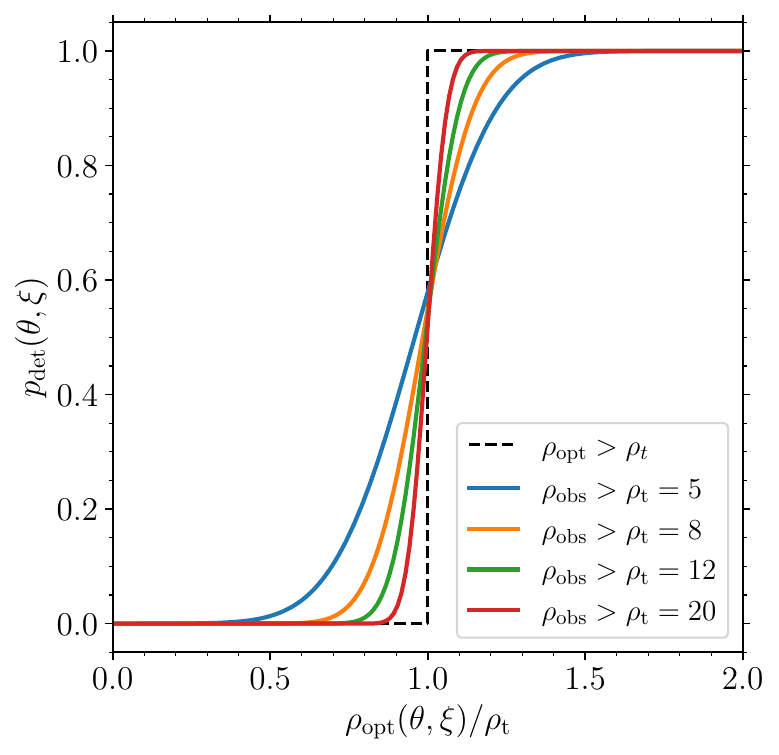}
\caption{Detectability $p_{\rm det}(\theta,\xi)$ of GW signals with given intrinsic and extrinsic parameters for a network of $N=3$ detectors as a function of the  optimal SNR $\rho_{\rm opt}(\theta,\xi)$. The black-dashed line is obtained by thresholding the optimal SNR itself, resulting in a step function centered on the threshold $\rho_{\rm t}$. Colored solid curves are obtained by thresholding the observed network SNR according to Eq.~(\ref{Qn2}) assuming various thresholds $\rho_{\rm t}$.} 
\label{marcumdistr}
\end{figure}

\begin{figure}\centering
\includegraphics[width=0.65\columnwidth]{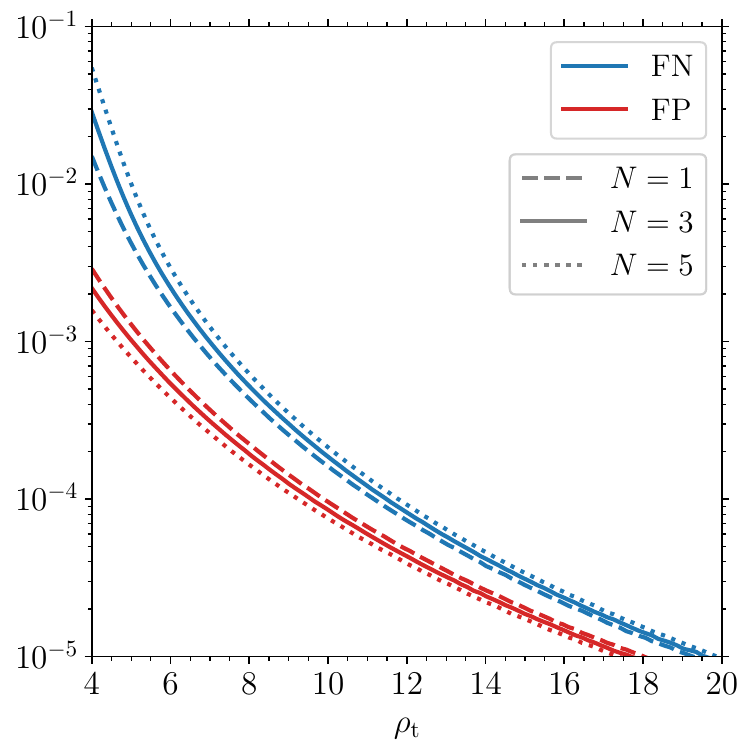}
\caption{Fraction of false negative (blue, FN) and false positive (red, FP) detections one would get by thresholding the optimal SNR $\rho_{\rm opt} $ instead of the matched filter SNR $\rho_{\rm obs}$. Both fractions decrease with the SNR threshold $\rho_{\rm t}$ and their difference increases increases with the number of detectors $N$. Dashed, solid, and dotted curves refers to networks of $N=1$, 3, and 5 interferometers, respectively.}
\label{falsepositives}
\end{figure}

In Fig.~\ref{marcumdistr}, we evaluate Eq.~(\ref{Qn2}) for the case of $N=3$ interferometers. As expected, thresholding the %
observed SNR smoothens the sharp step one would instead obtain when using the optimal SNR as detection statistics, allowing for some finite probability for events below (above) threshold to be detected (missed).  It is interesting to note that this effect is not symmetric: binaries with optimal network SNR that is below threshold are more likely to be detected than binaries with optimal network SNR above threshold are to be missed. That is, thresholding the optimal SNR $\rho_{\rm opt}$ instead of the matched filter SNR $\rho_{\rm obs}$ underestimates $p_{\rm det}$, hence overestimates the intrinsic merger rate. 
This effect is enhanced by the expected astrophysical SNR probability $p(\rho_{\rm opt})\propto \rho_{\rm opt}^{-4}$~\cite{2011CQGra..28l5023S}, which implies binaries are more likely to be found below than above threshold.  %

We further quantify this detection/non-detection asymmetry as follows, borrowing terminology from that of a classification problem \cite{2020sdmm.book.....I} where the actual outcome is given by $\rho_{\rm obs}>\rho_{\rm t}$ and the predicted outcome is given by the test $\rho_{\rm opt}>\rho_{\rm t}$. 
For a set of sources with SNRs distributed according to $p(\rho_{\rm opt})$, there are four mutually exclusive cases:
\begin{itemize}
\item $\rho_{\rm t}<\rho_{\rm obs}, \rho_{\rm opt}$ or ``true positives''. These events are flagged as detectable irrespectively of the thresholding strategy. The fraction of sources in this class is 
\begin{align}
{\rm TP} &= \int Q_{N/2}(\rho_{\rm opt} , \rho_{\rm t}) \;  \mathcal{I}(\rho_{\rm opt}>\rho_{\rm t}) \, p(\rho_{\rm opt}) \, \dd \rho_{\rm opt}
\end{align}
\item $\rho_{\rm obs}, \rho_{\rm opt}<\rho_{\rm t}$ or ``true negatives''. These events are flagged as non detectable irrespectively of the thresholding strategy.
The fraction of sources in this class is  
\begin{align}
{\rm TN} &= \int \left[1- Q_{N/2}(\rho_{\rm opt} , \rho_{\rm t})\right]\;  \mathcal{I}(\rho_{\rm opt}<\rho_{\rm t}) \, p(\rho_{\rm opt}) \, \dd \rho_{\rm opt}
\end{align}
\item $\rho_{\rm opt}< \rho_{\rm t}<\rho_{\rm obs}$ or ``false negatives''. These events are detectable but would be classified as non detectable if one neglects the SNR variance. The fraction of sources in this class is  
\begin{align}
{\rm FN} &= \int Q_{N/2}(\rho_{\rm opt} , \rho_{\rm t}) \;  \mathcal{I}(\rho_{\rm opt}<\rho_{\rm t}) \, p(\rho_{\rm opt}) \, \dd \rho_{\rm opt}
\end{align}
\item $\rho_{\rm obs}< \rho_{\rm t}<\rho_{\rm opt}$ or ``false positives''. These events are non detectable but would be classified as detectable if one neglects the SNR variance.  The fraction of sources in this class is  
\begin{align}
{\rm FP} &= \int \left[1- Q_{N/2}(\rho_{\rm opt} , \rho_{\rm t}) \right]\;  \mathcal{I}(\rho_{\rm opt}>\rho_{\rm t}) \, p(\rho_{\rm opt}) \, \dd \rho_{\rm opt}
\end{align}
\end{itemize}

Figure~\ref{falsepositives} shows the fractions FN and FP  as a function of the threshold $\rho_{\rm t}$ and the number of detectors $N$. We consider a population with $\rho_{\rm opt}\in [1,100]$ distributed according to $p(\rho_{\rm opt})\propto \rho_{\rm opt}^{-4}$. Both fractions go to zero as $\rho_{\rm t}$ increase, corresponding to the curves of Fig.~\ref{marcumdistr} approaching a step function: imposing a high detectability threshold makes it less likely for sources with a given optimal SNR to be scattered above or below threshold by a specific noise realization. The rate of false negatives is about an order of magnitude larger than that of false positives, confirming our point above. This difference increases with the number of detectors in the network, which is a consequence of Eq.~(\ref{Qdef}).

If needed, one can marginalize Eq.~(\ref{Qn2}) over the extrinsic parameters as in Eq.~(\ref{pdetthetadef}). Note however that the factorization of Eq.~(\ref{factoromega}) is not useful when considering multiple detectors because sources cannot be optimally oriented for all interferometers at the same time. %

\section{Applications to LIGO/Virgo}
\label{secapp}
In population studies, the quantity that enters the hierarchical likelihood~\cite{2019MNRAS.486.1086M,2022hgwa.bookE..45V}  is the expectation value (abusing notation once more) 
\begin{equation}
\label{pdetlambda}
p_{\rm det}(\lambda) = \int p_{\rm det}(\theta,\xi) p_{\rm pop}(\theta, \xi | \lambda) \dd \theta \dd \xi\,,
\end{equation}
where $\lambda$ indicates the population parameters and $p_{\rm pop}(\theta, \xi | \lambda)$ is the chosen population model. %
In case one is only trying to measure the population properties of the intrinsic parameters \cite{2023PhRvX..13a1048A}, then $p_{\rm pop}(\theta, \xi | \lambda) = p_{\rm pop}(\theta | \lambda) p(\xi)$
and 
$p_{\rm det}(\lambda) = \int p_{\rm det}(\theta) p_{\rm pop}(\theta | \lambda) \dd \theta$. In this section, we first compute population-averaged detectabilities on a controlled experiment and then use software injections in real LIGO noise.

\subsection{Toy population}

We estimate the impact of our findings on a toy population of GW events observable by the LIGO/Virgo network. We take a simple population $p_{\rm pop}(\theta, \xi | \lambda)$ where black-hole binaries have source-frame primary masses $m_1\in[5 M_\odot, 50 M_\odot]$  distributed according to $p(m_1)\propto m_1^{-2.3}$, source-frame secondary masses $m_2\in[5 M_\odot,m_1]$ distributed uniformly, redshifts $z\in[0,1]$ distributed uniformly in comoving volume and source-frame time 
$p(z)\propto (\dd V_c / \dd z)/ (1+z)$, spins magnitudes $\chi_{1,2}\in[0,1]$ distributed uniformly, and spin directions distributed isotropically. We assume a flat $\Lambda$CDM cosmological model with parameters from Ref.~\cite{2020A&A...641A...6P}. For the extrinsic parameters, we take uniform distributions in $\cos\iota$, $\cos\vartheta$, $\phi$, and $\psi$ as above. We consider a three-instrument network made of LIGO Livingston, LIGO Hanford, and Virgo at their design sensitivities~\cite{2018LRR....21....3A} and compute optimal SNRs $\rho_{\rm opt, i}$ using the %
the \textsc{IMRPhenomX} approximant~\cite{2021PhRvD.103j4056P}. For this example, our threshold is set to $\rho_{\rm t}=12$. %

\begin{figure}[t]
\centering
\includegraphics[width=0.65\columnwidth]{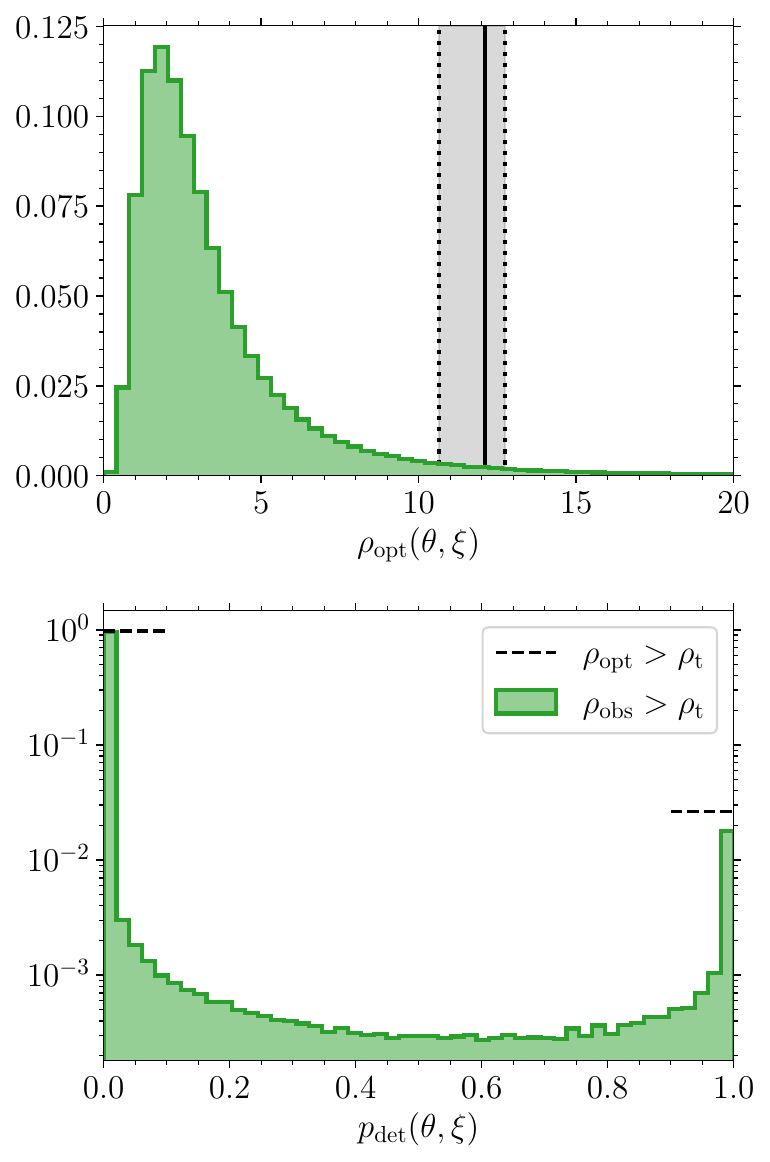}
\caption{Distribution of optimal SNR $\rho_{\rm opt}(\theta,\xi)$ (top) and detectabilities $p_{\rm det}(\theta,\xi)$ (bottom) for a toy population of black-hole binaries observable by the LIGO/Virgo network at design sensitivity, assuming a threshold $\rho_{\rm t}=12$. In the top panel, the green histogram shows the optimal SNRs. 
The vertical solid (dotted) black lines show the median (90\% inter-quantile range) of the probability of detection from Eq.~(\ref{Qn2}), indicating the region where the transition between non-detectability to the left and detectability to the right takes place. In the bottom panel, the green histogram shows the detectabilities obtained by thresholding the matched filter SNR $\rho_{\rm obs}$ and the two dashed black lines show the fractions of binaries that would be marked as detectable ($p_{\rm det}=1$) and non-detectable  ($p_{\rm det}=0$) if one were to instead threshold the optimal SNR~$\rho_{\rm opt}$. }
 \label{simplepop}
\end{figure}

Figure~\ref{simplepop} shows the resulting distributions of optimal network SNRs $\rho_{\rm opt}(\theta,\xi)$ and probabilities of detection $p_{\rm det}(\theta,\xi)$. A Monte Carlo estimate of the integral in Eq.~(\ref{pdetlambda}) computed over the entire population returns $p_{\rm det}(\lambda)\sim 0.027$ (we used $10^6$ samples, so the error on this number is $\ssim 10^{-3}$). %
If one only selects events with $|\rho_{\rm opt}(\theta,\xi) - \rho_{\rm t}|<1$ (i.e. those close to threshold), we find $p_{\rm det}(\lambda)\sim 0.496$. This should be compared with the analogous value $\ssim 0.435$ one would instead obtain with a simple cut on the optimal SNR. Marginalizing over noise realizations when estimating selection effects results in a higher $p_{\rm det}$ and mostly affects subpopulations of sources that are close to detection threshold.

\subsection{Pipeline injections} \label{pipeinj}

\begin{figure}[t]\centering
\includegraphics[width=0.9\columnwidth]{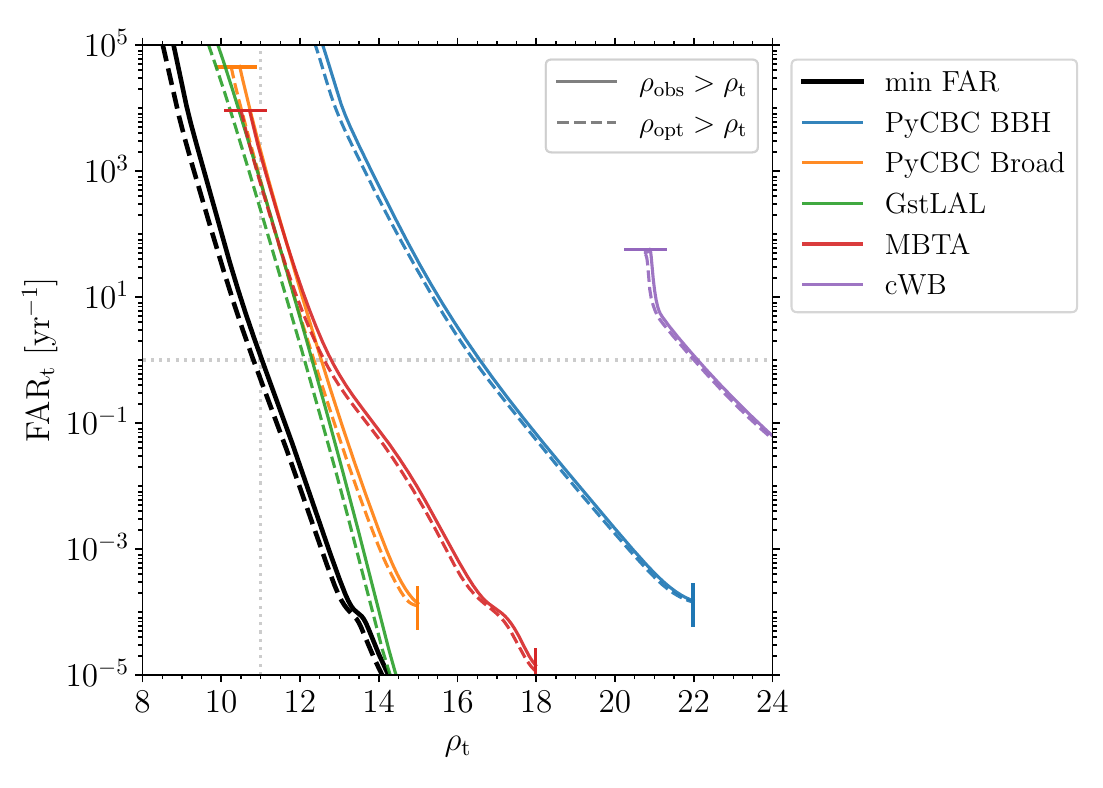}
\caption{Mapping between SNR threshold $\rho_{\rm t}$ and FAR threshold obtained by matching the population-averaged detectability $p_{\rm det}(\lambda)$. We use software injections into five different GW detection pipelines (colored curves) as well as the minimum FAR across all pipelines (black curves). Solid (dashed) curves are computed using the observed (optimal) SNR when thresholding for detectability.  Curves are restricted to the regime of validity of our analysis (see text). To guide the eye, the grey dotted lines indicate ${\rm FAR}_{\rm t} = 1/{\rm yr}$ and $\rho_{\rm t}=11$.
 } 
\label{getfar}
\end{figure}
It is important to remember that the SNR (either optimal or observed) provides approximate information on the GW detectability. The quantity returned by GW detection pipelines is the false-alarm rate (FAR), which is indeed the statistics in used state-of-the-art population analyses \cite{2023PhRvX..13a1048A} to both compile the list of events and estimate selection effects (other selection strategies are sometime used, see e.g. $p_{\rm astro}$ in Ref.~\cite{2023PhRvX..13d1039A}). Using the population average of Eq.~(\ref{pdetlambda}), we now present a calibration of the SNR threshold that enters our $p_{\rm det}$ approximation to reproduce the response of current detection pipelines as a function of their FARs.

We use software injections performed in real noise from the third LIGO/Virgo/KAGRA observing run \cite{inj}. %
They report FAR values for five different detection pipelines and LIGO-only (i.e. $N=2$) optimal SNRs for a fiducial population of sources (some of the injections also include Virgo; we neglect this difference and have verified it has a negligible impact on our results). Note their adopted population is different from that of our toy case above (see Ref.~\cite{inj} for details).  We use the injection dataset labeled as ``mixture,'' which is their recommended default. %
We consider all the injections provided, including cases where the FAR could not be quantified (and is reported as $\infty$ in the dataset).
Note the injections were performed uniformly in time, thus taking into account the duty cycle of the detectors. 

We compute the population-average detection probability by thresholding their FARs and match it with the population-average detection probability obtained with our $p_{\rm det}$ approach. Figure~\ref{getfar} shows the resulting mapping between the two quantities. We repeat our calculation for each of the five detection pipelines provided in the available dataset~\cite{inj} as well as by selecting the minimum FAR for each injected source. The latter is in line with the criterion used in Ref.~\cite{2023PhRvX..13a1048A} for selecting  compact binaries of astrophysical origin. 

This calculation is not reliable for SNRs that are $\lesssim 6$ because those injections were deemed ``hopeless'' to save computational time \cite{inj}. For those values of $\rho_{\rm t}$, there is a (potentially large) number of missing sources with optimal SNRs below threshold that could have been scattered above threshold by the SNR variance. 
We thus restrict our analysis to $\rho_{\rm t}\geq 8$, which is a few standard deviations away from the ``hopeless'' cut. %
Additionally, some pipelines have minimum and maximum FAR values they attempt to quantify. We estimated these limits from the provided datasets and truncated the curves in Fig.~\ref{getfar} accordingly. These limitations do not significantly impact the minimum FAR calculation across pipelines, at least in the region of interest shown in Fig.~\ref{getfar}.

Figure~\ref{getfar} shows that %
 selection effects computed using a minimum  FAR threshold of $\ssim 1$ yr$^{-1}$ are reproduced by an SNR-based cut with threshold {$\rho_{\rm t}\simeq 11$}. We find once more that the bias induced by thresholding $\rho_{\rm opt}$ instead of $\rho_{\rm obs}$ is of a few \% and underestimates $p_{\rm det}$ across the entire range: that is, the $\rho_{\rm t}$ threshold for a given FAR obtained when marginalizing over noise realization is larger compared to the case where noise is neglected. The difference between the two treatments is smaller than the typical difference between the various detection pipelines (and indeed %
current LIGO/Virgo selection criteria combine information from multiple pipelines). The population used in Ref.~\cite{inj} is deliberately broad, while our results above indicate that the subpopulation of sources that are close to threshold %
 will be affected more significantly by the SNR variance. Confirming this expectation with dedicated pipeline injections is left to  future work. %

As shown in Fig.~\ref{getfar}, we empirically find that the mapping between $\rho_{\rm t}$ and the minimum FAR threshold across the available pipelines is well described by a power law:
\begin{align}
\log_{10} \left(\frac{{\rm FAR}_{\rm t}}{{\rm yr}^{-1}}\right)= p_1\, \rho_{\rm t} +p_0.
\label{farfit}
\end{align}
A simple least-square regression returns {$p_1= -1.80$} and {$p_0= 20.1$} when thresholding using $\rho_{\rm obs}$ (i.e. this is a fit to the solid black curve in Fig.~\ref{getfar}),  and  {$p_1= -1.72$} and {$p_0= 18.7$} when thresholding using $\rho_{\rm opt}$ (i.e. this is a fit to the dashed black curve in Fig.~\ref{getfar}). %
These fits can be used to quickly filter synthetic distributions according to the desired purity of the resulting GW simulated catalog, though with the important caveat that this relationship was calibrated on a specific population of sources.
We stress that Fig.~\ref{getfar} and Eq.~(\ref{farfit}) present a calibration between thresholds, 
not a mapping of SNR to FAR for a given trigger and pipeline.

\section{Summary}
\label{secend}

We summarized and extended the most common treatment of GW selection effects, namely that of thresholding the SNR. We focused in particular on the marginalization over extrinsic parameters and noise realizations, considering both single and multiple detectors. 

When modeling the filter imposed to observations by the detectors, 
the simplest strategy %
is to consider a source ``detectable'' if its optimal SNR $\rho_{\rm opt}$ is sufficiently large~\cite{1993PhRvD..47.2198F}. This approach does not take into account the SNR variance induced by noise realizations. As presented here, incorporating such an effect is straightforward and results in a closed-form expression of $p_{\rm det}$. All one needs to do is substitute the sharp step
\begin{equation}
p_{\rm det}(\theta,\xi) = \mathcal{I}[\rho_{\rm opt} (\theta,\xi) > \rho_{\rm t}]
\end{equation}
with the smooth transition
\begin{equation}
p_{\rm det}(\theta,\xi) = Q_{N/2}\big[\rho_{\rm opt} (\theta,\xi), \rho_{\rm t} \big]\,.
\end{equation}

Including noise realizations when estimating $p_{\rm det}$  takes care of the conceptual %
point recently raised in Ref.~\cite{2024ApJ...962..169E}. There the authors argue  that a consistent detectability estimate should never make use of the true signal parameters, which are not accessible, but only of the data, and these are inevitably affected by noise.

Our expression is appealing for its simplicity but still approximate. Notable simplifications include (i) assuming that noise is stationary and Gaussian, which is never perfectly the case, (ii)~thresholding events using the SNR and not the  FAR, and (iii)  neglecting
errors due to the finite size of our template banks~\cite{2023PhRvD.108d3011E}. Further improvements  include considering the phase-maximized SNR, which can also be written down analytically using special functions~\cite{2017PhRvD..96d4005C}.   %

 We find that a SNR threshold of {$\ssim 11$}  reproduces the %
 selection criterion used in current analyses (FAR < 1 yr$^{-1}$ for binary black holes \cite{2023PhRvX..13a1048A}) and that the inclusion of noise realizations increases the average detection probability $p_{\rm det}$ by a few \%. While nominally modest, this effect becomes increasingly important as the number of detections grows because systematic errors related to selection effects scale faster than linearly  with the number of events in the catalog \cite{2023MNRAS.526.3495T}. Furthermore, the projected bias is not uniform across the parameter space and disproportionally affects sources at the edge of detectability. For instance, this will be relevant to analyses targeting compact binaries at high redshift \cite{2020ApJ...891L..31F,2023PhRvD.107j1302M,2024ApJ...962..169E} and attempting to discriminate their origin as either astrophysical or cosmological \cite{2022ApJ...931L..12N,2023arXiv231018158F}.

Accurate modeling of selection effects 
is prominent in both (i) GW population studies, where selection effects enter the hierarchical likelihood, and (ii) the development of astrophysical predictions, where outputs of population-synthesis codes are post-processed to obtain detectable distributions.
We hope the ``collection of recipes'' we presented here will provide a useful companion to researchers working in either of these two contexts, facilitating the treatment of selection biases in GW astronomy.

\section*{Acknowledgments}

In memory of Chris Belczynski. I first got interested in selection effects while post-processing your simulations. I'm sure you're on a beautiful mountain; farewell. (D.G.) \medskip \vspace{-0.4cm}

\noindent We thank Matt Mould, Costantino Pacilio, Michele Mancarella, Viola De Renzis, Francesco Iacovelli, Nick Loutrel, Chris Moore, Riccardo Buscicchio, Reed Essick and the participants of the \textit{``Gravitational-wave populations: what's next?''} conference (Milan, 2023) for discussions. We thank the referees for their inputs on Sec.~\ref{pipeinj}. 
D.G. and M.B. are supported by ERC Starting Grant No.~945155--GWmining, 
Cariplo Foundation Grant No.~2021-0555, MUR PRIN Grant No.~2022-Z9X4XS, 
and the ICSC National Research Centre funded by NextGenerationEU. 
D.G. is supported by MSCA Fellowships No.~101064542--StochRewind and No.~101149270--ProtoBH.
Computational work was performed at CINECA with allocations 
through~INFN~and~Bicocca.

\section*{ORCID IDs}
Davide Gerosa \orcidlink{0000-0002-0933-3579} \href{https://orcid.org/0000-0002-0933-3579}{https://orcid.org/0000-0002-0933-3579} \\
Malvina Bellotti \orcidlink{0009-0003-1050-9273} \href{https://orcid.org/0009-0003-1050-9273}{https://orcid.org/0009-0003-1050-9273}

\appendix
\section{Marcum $Q$-functions with Python}
\label{python}

The numerical implementation of the Marcum $Q$-function for the Python programming language is somewhat hidden in the popular module \textsc{scipy} \cite{2020NatMe..17..261V}. In particular, object \textsc{scipy.stats.ncx2} provides  tools to characterize the non-central $\chi^2$ distribution (which is different than the non-central $\chi$ distribution used in Sec.~\ref{multdetmatched}). The probability density function of a random variate $x$ distributed according to a non-central $\chi^2$ distribution  with $k$ degrees of freedom and non-centrality parameter $\lambda$ is 
\begin{equation}
p(x)= {\frac {1}{2}}\left({\frac {x}{\lambda }}\right)^{k/4-1/2} \exp\left(-\frac{x+\lambda}{2}\right)I_{k/2-1}\left({\sqrt {\lambda x}}\right)\,.
\end{equation}
Integrating this, one obtains the complementary cumulative distribution function (or survival function) %
\begin{equation}
p(x>x_0) = Q_{k/2}\left(\sqrt{\lambda}, \sqrt{x_0}\right)\,.
\end{equation}
The expression $Q_\nu(a,b)$ from Eq.~(\ref{Qdef}) can therefore be evaluated as the survival function of a non-central $\chi^2$ distribution with $x= b^2$, $k=2\nu$ and $\lambda =a^2$. In Python, this is 
\begin{verbatim}
    def marcumq(nu,a,b):
        return scipy.stats.ncx2.sf(b**2, 2*nu, a**2)
\end{verbatim}

We have implemented this  function in a standalone package named \textsc{marcumq}~\cite{repo}. This can be installed with
\begin{verbatim}
    pip install marcumq
\end{verbatim}
and used with e.g.
\begin{verbatim}
    import marcumq
    marcumq.marcumq(nu,a,b)
\end{verbatim}
We have tested our implementation against those provided in the symbolic manipulation tools \textsc{sympy} and \textsc{mathematica}.

\section*{References}
\bibliographystyle{iopart-num_leo}
\bibliography{pdetvar}

\providecommand{\newblock}{}
\begin{thebibliography}{10}
\expandafter\ifx\csname url\endcsname\relax
  \def\url#1{{\tt #1}}\fi
\expandafter\ifx\csname urlprefix\endcsname\relax\def\urlprefix{URL }\fi
\providecommand{\eprint}[2][]{\href{http://arxiv.org/abs/#2}{arXiv:#2}}

\bibitem{2019MNRAS.486.1086M}
{Mandel} I, {Farr} W~M and {Gair} J~R 2019
  \href{http://dx.doi.org/10.1093/mnras/stz896}{ {\em Mon. Not. R. Astron.
  Soc.\/} {\bf 486} 1086--1093 } [\eprint{1809.02063}]

\bibitem{2022hgwa.bookE..45V}
{Vitale} S, {Gerosa} D, {Farr} W~M and {Taylor} S~R 2022
  \href{http://dx.doi.org/10.1007/978-981-15-4702-7_45-1}{ {Inferring the
  Properties of a Population of Compact Binaries in Presence of Selection
  Effects} } {\em Handbook of Gravitational Wave Astronomy\/} (Springer) p~45

\bibitem{2023MNRAS.526.3495T}
{Talbot} C and {Golomb} J 2023 \href{http://dx.doi.org/10.1093/mnras/stad2968}{
  {\em Mon. Not. R. Astron. Soc.\/} {\bf 526} 3495--3503 }
  [\eprint{2304.06138}]

\bibitem{2023PhRvX..13a1048A}
{Abbott} R {\em et~al.\/} 2023
  \href{http://dx.doi.org/10.1103/PhysRevX.13.011048}{ {\em Phys. Rev. X\/}
  {\bf 13} 011048 } [\eprint{2111.03634}]

\bibitem{2018CQGra..35n5009T}
{Tiwari} V 2018 \href{http://dx.doi.org/10.1088/1361-6382/aac89d}{ {\em Class.
  Quantum Grav.\/} {\bf 35} 145009 } [\eprint{1712.00482}]

\bibitem{calibratedVT}
{Wysocki} D and {O'Shaughnessy} R 2018 {\em
  \href{https://dcc.ligo.org/LIGO-T1800427/public}{LIGO Document T1800427}\/}

\bibitem{2019RNAAS...3...66F}
{Farr} W~M 2019 \href{http://dx.doi.org/10.3847/2515-5172/ab1d5f}{ {\em Res.
  Notes AAS\/} {\bf 3} 66 } [\eprint{1904.10879}]

\bibitem{1993PhRvD..47.2198F}
{Finn} L~S and {Chernoff} D~F 1993
  \href{http://dx.doi.org/10.1103/PhysRevD.47.2198}{ {\em Phys. Rev. D\/} {\bf
  47} 2198--2219 } [\eprint{gr-qc/9301003}]

\bibitem{2018LRR....21....3A}
{Abbott} B~P {\em et~al.\/} 2018
  \href{http://dx.doi.org/10.1007/s41114-018-0012-9}{ {\em Living Rev.
  Relativ.\/} {\bf 21} 3 } [\eprint{1304.0670}]

\bibitem{2023PhRvD.108d3011E}
{Essick} R 2023 \href{http://dx.doi.org/10.1103/PhysRevD.108.043011}{ {\em
  Phys. Rev. D\/} {\bf 108} 043011 } [\eprint{2307.02765}]

\bibitem{2021ApJ...922..258V}
{Veske} D, {Bartos} I, {M{\'a}rka} Z and {M{\'a}rka} S 2021
  \href{http://dx.doi.org/10.3847/1538-4357/ac27ac}{ {\em Astrophys. J.\/} {\bf
  922} 258 } [\eprint{2105.13983}]

\bibitem{2020PhRvD.102j3020G}
{Gerosa} D, {Pratten} G and {Vecchio} A 2020
  \href{http://dx.doi.org/10.1103/PhysRevD.102.103020}{ {\em Phys. Rev. D\/}
  {\bf 102} 103020 } [\eprint{2007.06585}]

\bibitem{2022ApJ...927...76T}
{Talbot} C and {Thrane} E 2022
  \href{http://dx.doi.org/10.3847/1538-4357/ac4bc0}{ {\em Astrophys. J.\/} {\bf
  927} 76 }

\bibitem{2024PhRvD.109f3013M}
{Mould} M, {Moore} C~J and {Gerosa} D 2024
  \href{http://dx.doi.org/10.1103/PhysRevD.109.063013}{ {\em Phys. Rev. D\/}
  {\bf 109} 063013 } [\eprint{2311.12117}]

\bibitem{Marcum1948AST}
Marcum J 1948
  \href{http://dx.doi.org/https://api.semanticscholar.org/CorpusID:12688986}{
  {\em IRE Trans. Inf. Theory\/} {\bf 6} 59--267 }

\bibitem{Shnidman1989TheCO}
Shnidman D~A 1989
  \href{http://dx.doi.org/https://api.semanticscholar.org/CorpusID:39701757}{
  {\em IEEE Trans. Inf. Theory\/} {\bf 35} 389--400 }

\bibitem{2013arXiv1311.0681G}
{Gil} A, {Segura} J and {Temme} N~M 2013
  \href{http://dx.doi.org/10.1145/2591004}{ {\em ACM Trans. Math. Softw.\/}
  {\bf 40} 1 } [\eprint{1311.0681}]

\bibitem{2020ApJ...891L..31F}
{Fishbach} M, {Farr} W~M and {Holz} D~E 2020
  \href{http://dx.doi.org/10.3847/2041-8213/ab77c9}{ {\em Astrophys. J.
  Lett.\/} {\bf 891} L31 } [\eprint{1911.05882}]

\bibitem{2023PhRvD.107j1302M}
{Mancarella} M, {Iacovelli} F and {Gerosa} D 2023
  \href{http://dx.doi.org/10.1103/PhysRevD.107.L101302}{ {\em Phys. Rev. D\/}
  {\bf 107} L101302 } [\eprint{2303.16323}]

\bibitem{2024ApJ...962..169E}
{Essick} R and {Fishbach} M 2024
  \href{http://dx.doi.org/10.3847/1538-4357/ad1604}{ {\em Astrophys. J.\/} {\bf
  962} 169 } [\eprint{2310.02017}]

\bibitem{Maggiore:2007ulw}
Maggiore M 2007
  \href{http://dx.doi.org/10.1093/acprof:oso/9780198570745.001.0001}{ {\em
  {Gravitational Waves. Vol. 1: Theory and Experiments}\/} } (Oxford)

\bibitem{2011gwpa.book.....C}
{Creighton} J and {Anderson} W 2011 {\em {Gravitational-Wave Physics and
  Astronomy: An Introduction to Theory, Experiment and Data Analysis.}\/}
  (Wiley)

\bibitem{2019PASA...36...10T}
{Thrane} E and {Talbot} C 2019 \href{http://dx.doi.org/10.1017/pasa.2019.2}{
  {\em Publ. Astron. Soc. Aust.\/} {\bf 36} e010 } [\eprint{1809.02293}]

\bibitem{2011CQGra..28l5023S}
{Schutz} B~F 2011 \href{http://dx.doi.org/10.1088/0264-9381/28/12/125023}{ {\em
  Class. Quantum Grav.\/} {\bf 28} 125023 } [\eprint{1102.5421}]

\bibitem{2016PhRvD..94l1501V}
{Vitale} S 2016 \href{http://dx.doi.org/10.1103/PhysRevD.94.121501}{ {\em Phys.
  Rev. D\/} {\bf 94} 121501 } [\eprint{1610.06914}]

\bibitem{Park1961MomentsOT}
Park J~H 1961
  \href{http://dx.doi.org/https://api.semanticscholar.org/CorpusID:125755710}{
  {\em Q. Appl. Math.\/} {\bf 19} 45--49 }

\bibitem{1972hmfw.book.....A}
{Abramowitz} M and {Stegun} I~A 1972 {\em {Handbook of Mathematical
  Functions}\/} (Dover)

\bibitem{2015PhRvD..91l4062G}
{Gair} J~R and {Moore} C~J 2015
  \href{http://dx.doi.org/10.1103/PhysRevD.91.124062}{ {\em Phys. Rev. D\/}
  {\bf 91} 124062 } [\eprint{1504.02767}]

\bibitem{2012LRR....15....4J}
{Jaranowski} P and {Kr{\'o}lak} A 2012
  \href{http://dx.doi.org/10.12942/lrr-2012-4}{ {\em Living Rev. Relativ.\/}
  {\bf 15} 4 } [\eprint{0711.1115}]

\bibitem{repo}
{Gerosa} D 2023 {\em \href{https://doi.org/10.5281/zenodo.10071541}{Zenodo
  10071541}\/}
  \href{https://github.com/dgerosa/marcumq}{github.com/dgerosa/marcumq}

\bibitem{2020sdmm.book.....I}
{Ivezi{\'c}} {\v{Z}}, {Connolly} A~J, {VanderPlas} J~T and {Gray} A 2020
  \href{http://dx.doi.org/10.1515/9780691197050}{ {\em {Statistics, Data
  Mining, and Machine Learning in Astronomy. A Practical Python Guide for the
  Analysis of Survey Data}\/} } (Princeton)

\bibitem{2020A&A...641A...6P}
{Aghanim} N {\em et~al.\/} 2020
  \href{http://dx.doi.org/10.1051/0004-6361/201833910}{ {\em Astron.
  Astrophys.\/} {\bf 641} A6 } [\eprint{1807.06209}]

\bibitem{2021PhRvD.103j4056P}
{Pratten} G, {Garc{\'\i}a-Quir{\'o}s} C, {Colleoni} M, {Ramos-Buades} A,
  {Estell{\'e}s} H, {Mateu-Lucena} M, {Jaume} R, {Haney} M, {Keitel} D,
  {Thompson} J~E and {Husa} S 2021
  \href{http://dx.doi.org/10.1103/PhysRevD.103.104056}{ {\em Phys. Rev. D\/}
  {\bf 103} 104056 } [\eprint{2004.06503}]

\bibitem{2023PhRvX..13d1039A}
{Abbott} R {\em et~al.\/} 2023
  \href{http://dx.doi.org/10.1103/PhysRevX.13.041039}{ {\em Phys. Rev. X\/}
  {\bf 13} 041039 } [\eprint{2111.03606}]

\bibitem{inj}
{LIGO, Virgo, and KAGRA Collaboration} 2023 {\em
  \href{https://zenodo.org/records/7890437}{Zenodo 7890437}\/}

\bibitem{2017PhRvD..96d4005C}
{Chua} A~J~K, {Moore} C~J and {Gair} J~R 2017
  \href{http://dx.doi.org/10.1103/PhysRevD.96.044005}{ {\em Phys. Rev. D\/}
  {\bf 96} 044005 } [\eprint{1705.04259}]

\bibitem{2022ApJ...931L..12N}
{Ng} K~K~Y, {Chen} S, {Goncharov} B, {Dupletsa} U, {Borhanian} S, {Branchesi}
  M, {Harms} J, {Maggiore} M, {Sathyaprakash} B~S and {Vitale} S 2022
  \href{http://dx.doi.org/10.3847/2041-8213/ac6bea}{ {\em Astrophys. J.
  Lett.\/} {\bf 931} L12 } [\eprint{2108.07276}]

\bibitem{2023arXiv231018158F}
{Fairhurst} S, {Mills} C, {Colpi} M, {Schneider} R, {Sesana} A, {Trinca} A and
  {Valiante} R 2024 \href{http://dx.doi.org/10.1093/mnras/stae443}{ {\em Mon.
  Not. R. Astron. Soc.\/} } [\eprint{2310.18158}]

\bibitem{2020NatMe..17..261V}
{Virtanen} P {\em et~al.\/} 2020
  \href{http://dx.doi.org/10.1038/s41592-019-0686-2}{ {\em Nat. Methods\/} {\bf
  17} 261--272 } [\eprint{1907.10121}]

\end{thebibliography}

\end{document}